\documentclass[prl,aps,showpacs,twocolumn,floatfix]{revtex4}
\usepackage{epsfig} \usepackage{graphics} \usepackage{bm}
\usepackage{amssymb}
\usepackage{graphicx}
\begin{document}

\preprint{Lebed-Rapids}

\title{Quantum limit in a quasi-one-dimensional conductor in a high tilted
magnetic field}

\author{A.G. Lebed$^*$}

\affiliation{Department of Physics, University of Arizona, 1118 E.
4-th Street, Tucson, AZ 85721, USA}

\begin{abstract}
Recently, we have suggested Fermi-liquid - non-Fermi-liquid
angular crossovers which may exist in quasi-one-dimensional (Q1D)
conductors in high tilted magnetic fields [see A.G. Lebed, Phys.
Rev. Lett. $\textbf{115}$, 157001 (2015).] All calculations in the
Letter, were done by using the quasi-classical Peierls substitution
method, whose applicability in high magnetic fields was
questionable. Here, we solve a fully quantum mechanical problem
and show that the main qualitative conclusions of the above
mentioned Letter are correct. In particular, we show that in high
magnetic fields, applied along one of the two main crystallographic axis, we
have 2D electron spectrum, whereas, for directions of high magnetic
fields far from the axes, we have 1D electron spectrum. The later
is known to promote non-Fermi-liquid properties. As a result, we
expect the existence of Fermi-liquid - non-Fermi-liquid angular
crossovers or phase transitions. Electronic parameters of Q1D
conductor (Per)$_2$Pt(mnt)$_2$ show that such transitions can
appear in feasible high magnetic fields of the order of $H \simeq
20-25 \ T$.
\end{abstract}


\maketitle

Fermi-liquid theory has been successful in explanations of very
unusual magnetic properties in quasi-one-dimensional (Q1D)
conductors [1]. The first historically discovered such properties,
the so-called Field-Induced Spin-Density-Waves and related 3D
quantum Hall effect in Q1D conductors (TMTSF)$_2$X (X=ClO$_4$,
PF$_6$, NO$_3$, etc.) [2,3], were explained [1,4-10] by studying
peculiarities of semiclassical electron motion along open Q1D
sheets of the Fermi surface and electron-hole interactions under
such conditions. New concept, $3D \rightarrow 2D$ dimensional
crossover, was introduced [4] in the framework of Fermi-liquid
approach. Extensions of this concept to different $3D \rightarrow
1D \rightarrow 2D$ crossovers in a tilted magnetic field allowed
to explain such phenomena as the so-called Lebed's magic angles
(LMA) and Lee-Naughton-Lebed's oscillations (see, for example,
Ref.[11] and references therein). It is important that all the
above mentioned physical phenomena happen in low enough magnetic
fields, where the "sizes" of electron trajectories are less than
inter-chain and inter-plane distances. Despite the above-described
success, non-Fermi-liquid properties have been also studied both
theoretically [12-15] and experimentally. For instance, there
exist experimental claims that Fermi liquid cannot explain
adequately the LMA phenomena in the Nernst [16-19] and Hall [20]
effects.

An important step in studying magnetic properties of layered Q1D
compounds was made in Ref.[21], where it was suggested that
some other types of $3D \rightarrow 2D$ crossovers - quantum
dimensional crossovers - can occur in such compounds in high
magnetic fields. It was shown [21] that, at high magnetic fields,
the typical "sizes" of electron trajectories can become less than
inter-layered distances in layered Q1D conductors, which results in
the appearance of such unusual phenomenon as the Reentrant
Superconductivity [21-23]. Different types of quantum $3D
\rightarrow 2D$ and $3D \rightarrow 1D$ dimensional crossovers
have been recently studied theoretically [24-27]. In particular,
in Ref.[26], it was shown that, when strong magnetic field was
applied far enough from the main crystallographic axes of a Q1D
conductor, we have almost $1D$ electron spectrum and can expect
non-Fermi-liquid behavior, whereas, for the magnetic field
directions close to the main crystallographic axis, we have 2D
electron spectrum and, thus, Fermi-liquid properties have to restore. On
this basis, in Ref.[26], Fermi-liquid - non-Fermi-liquid angular
crossovers were suggested to occur in a Q1D conductor in a high
magnetic field and the critical magnetic field was estimated as $H
\simeq 25 \ T$ in layered Q1D conductor (Per)$_2$Pt(mnt)$_2$.

We pay attention that all previous considerations of the quantum
dimensional crossovers [21-27] are based on quasi-classical
version of the so-called Peierls substitution method [4,1]. The
goal of our paper is to solve a fully quantum mechanical problem
for a Q1D geometry of electron spectrum in a perpendicular
magnetic field. We show that at high magnetic fields electron wave
functions indeed localize on the 1D conducting chains, unless the
magnetic field is directed along one of the main crystallographic
axes. If a magnetic field is directed along one of the axes, it is
shown that electron wave functions are localized only on 2D
planes. This property of electron wave functions confirms the
hypothesis of Ref.[26] about possible angular Fermi-liquid -
non-Fermi-liquid crossovers (or phase transitions).

At first, let us consider a $3D$ isotropic electron spectrum,
\begin{equation}
\epsilon(p_x,p_y,p_z)= \frac{p^2_x}{2m} + \frac{p^2_y+p_x^2}{2m},
\end{equation}
where $\frac{p^2_x}{2m}$ is electron energy along the conducting
chains, in the following magnetic field, inclined perpendicular to
the chains:
\begin{eqnarray}
&{\bf H}=(0, H \sin \alpha , H \cos \alpha),
\nonumber\\
&{\bf A}=(Hz \sin \alpha - Hy \cos \alpha, 0, 0).
\end{eqnarray}
It is important that electron motion along the chains is free and
characterized by the large energies, $\frac{p^2_x}{2m} \sim
\epsilon_F$, where $\epsilon_F$ is the Fermi energy. Therefore,
the Peierls substitution method for momentum $p_x$ in Eq.(1) is an
exact procedure, $p_x \rightarrow -i \frac{\partial}{\partial x} -
(e/c)A_x$, where $\hbar \equiv 1$. As a result, we obtain for
kinetic energy in magnetic field (2) the following operator:
\begin{eqnarray}
&\hat \epsilon(x,y,z)= \frac{1}{2m}
\biggl[ \biggl( -i \frac{\partial}{\partial x} + \frac{eHy \cos \alpha}{c}
-\frac{eHz \sin \alpha}{c} \biggl)^2
\nonumber\\
&+\biggl( -i \frac{\partial}{\partial y} \biggl)^2
+ \biggl( -i \frac{\partial}{\partial y} \biggl)^2 \biggl] .
\end{eqnarray}

As seen from Eq.(3), the magnetic field (2) does not disturb
electron motion along the chains, where electrons are
characterized by conserved momenta close to the Fermi momentum,
$p_F$. Therefore, we represent electron wave functions in the
following way:
\begin{equation}
\Psi_{\epsilon}(x,y,z) = \exp(ip_x x) \Psi_\epsilon(y,z) , \ \ \
p_x \simeq p_F.
\end{equation}
Substitution of the wave function (4) into the kinetic energy operator
(3) results in
\begin{eqnarray}
&\hat \epsilon(y,z)= \frac{1}{2m}
\biggl[ \biggl( p_F + \frac{eHy \cos \alpha}{c}
-\frac{eHz \sin \alpha}{c} \biggl)^2
\nonumber\\
&+\biggl( -i \frac{\partial}{\partial y} \biggl)^2
+ \biggl( -i \frac{\partial}{\partial y} \biggl)^2 \biggl] .
\end{eqnarray}
Note that below we keep in electron energy (5) only terms of the
order of $\omega_b(\alpha)$ and $\omega_c(\alpha)$ and disregard
terms of the order of $\omega^2_b(\alpha)/\epsilon_F \ll
\omega_b(\alpha)$ and $\omega^2_c(\alpha)/\epsilon_F \ll
\omega_c(\alpha)$, where
\begin{equation}
\omega_b(\alpha)=\frac{eHv_Fb^*\cos \alpha}{c}, \ \
\omega_c(\alpha)=\frac{eHv_Fc^* \sin \alpha}{c}
\end{equation}
are the so-called cyclotron frequencies [26].
 Here, $b^*$ and
$c^*$ are the crystalline lattice parameters along ${\bf y}$ and
${\bf z}$ axes, correspondingly; $v_F=p_F/m$ is the Fermi velocity
along the conducting ${\bf x}$ axis. Therefore, we can linearize
kinetic energy operator (5) with respect to the frequencies,
$\omega_b(\alpha)$ and $\omega_c(\alpha)$:
\begin{eqnarray}
&\frac{1}{2m} \biggl(p_F + \frac{eHy \cos \alpha}{c}
-\frac{eHx \sin \alpha}{c} \biggl)^2
\nonumber\\
&=\frac{1}{2m} \biggl[p_F + \frac{\omega_b(\alpha)}{v_F} \frac{y}{b^*}
- \frac{\omega_c(\alpha)}{v_F} \frac{z}{c^*} \biggl]^2
\nonumber\\
&\simeq \epsilon_F + \omega_b(\alpha) \frac{y}{b^*}
- \omega_c(\alpha) \frac{z}{c^*} .
\end{eqnarray}

If electron potential energy in (${\bf y},{\bf z}$) plane,
perpendicular to the conducting axis, is $V(y,z)$, then the
corresponding Schr\"{o}dinger equation for electron wave functions
can be written as
\begin{eqnarray}
\biggl[ -\frac{1}{2m} \biggl( \frac{\partial^2}{\partial y^2}
+ \frac{\partial^2}{\partial z^2} \biggl)
+ \omega_b(\alpha) \frac{y}{b^*} - \omega_c(\alpha) \frac{z}{c^*}
\nonumber\\
+ V(y,z) \biggl] \Psi_{\tilde \epsilon}(y,z)
= \tilde \epsilon \Psi_{\tilde \epsilon}(y,z) ,
\end{eqnarray}
where $\tilde \epsilon = \epsilon - \epsilon_F$. Below, we
consider the following model for the in-plane potential energy,
\begin{equation}
V(y,z) = - \frac{\kappa_1}{m} \sum_{n_1 = - \infty}^{+ \infty}
\delta(y - n_1 b^*) - \frac{\kappa_2}{m} \sum_{n_2 = - \infty}^{+
\infty} \delta(z-n_2 c^*) ,
\end{equation}
where $\delta(y)$ and $\delta(z)$ are the 1D Dirac
delta-functions. In this case, as we show below, a tight-binding
variant of the Scr\"{o}dinger equation in a magnetic field (8),(9)
becomes exactly solvable. In particular, the total 2D wave
function in Eq.(8) can be factorized in our case:
\begin{equation}
\Psi(y,z) = \Psi_{\tilde \epsilon}(y) \Psi_{\tilde \epsilon}(z) .
\end{equation}
For 1D wave functions (10), it is easy to obtain the following
Schr\"{o}dinger equations:
\begin{eqnarray}
&\biggl[ -\frac{1}{2m} \biggl( \frac{d^2}{d y^2} \biggl) +
\omega_b(\alpha) \frac{y}{b^*} - \frac{\kappa_1}{m} \sum_{n_1 = -
\infty}^{+ \infty} \delta(y - n_1 b^*) \biggl]
\nonumber\\
&\times \Psi_{\tilde \epsilon_1}(y) = \tilde \epsilon_1
\Psi_{\tilde \epsilon_1}(y)
\end{eqnarray}
and
\begin{eqnarray}
&\biggl[ -\frac{1}{2m} \biggl( \frac{d^2}{d z^2} \biggl) -
\omega_c(\alpha) \frac{z}{b^*} - \frac{\kappa_2}{m} \sum_{n_2 = -
\infty}^{+ \infty} \delta(z - n_2 c^*) \biggl]
\nonumber\\
&\times \Psi_{\tilde \epsilon_2}(z) = \tilde \epsilon_2
\Psi_{\tilde \epsilon_2}(z),
\end{eqnarray}
where $\tilde \epsilon = \tilde \epsilon_1 + \tilde \epsilon_2$.

Let us first consider Eq.(11) for $y$ coordinate. To solve
Eq.(11), we use the so-called tight-binding approximation, where
the wave function can be expressed as
\begin{equation}
\Psi_{\tilde \epsilon_1}(y) = \sum_{m=-\infty}^{+\infty} A_m
(H,\alpha) \ \phi_{\epsilon_{01}}(y - m b^*).
\end{equation}
Here, the wave function,
\begin{equation}
\phi_{\epsilon_{01}}(y) = \sqrt{\kappa_1} \exp(-\kappa_1 |y|), \ \
\epsilon_{01} = - \frac{\kappa^2_1}{2m},
\end{equation}
is solution of the following equation:
\begin{equation}
\biggl[ -\frac{1}{2m} \biggl( \frac{d^2}{d y^2} \biggl) -
\frac{\kappa_1}{m} \ \delta(y) \biggl] \phi_{\epsilon_{01}}(y)=
\epsilon_{01} \phi_{\epsilon_{01}}(y).
\end{equation}
By using the tight-binding approximation, it is possible to show
that the amplitudes $A_m (H, \alpha)$ in Eq.(13) satisfy the
following equation:
\begin{eqnarray}
&A_m(H,\alpha)[\tilde \epsilon_1 - \epsilon_{01} - m \
\omega_b(\alpha)]
\nonumber\\
&+ A_{m+1}(H, \alpha) \ t_1 + A_{m-1}(H, \alpha) \ t_1 = 0 ,
\end{eqnarray}
where
\begin{equation}
t_1 =
\frac{\kappa_1}{m}\phi_{\epsilon_{01}}(0)\phi_{\epsilon_{01}}(\pm
b^*)= \frac{\kappa^2_1}{m} \exp(-\kappa_1 b^*).
\end{equation}

Now, we consider Eq.(12) for $z$ coordinate. As it was done above,
to solve Eq.(12), we use the tight-binding approximation, where
the wave function (12) can be written as
\begin{equation}
\Psi_{\tilde \epsilon_2}(z) = \sum_{l=-\infty}^{+\infty} B_l
(H,\alpha) \ \phi_{\epsilon_{02}}(z - l c^*),
\end{equation}
where the wave function,
\begin{equation}
\phi_{\epsilon_{02}}(z) = \sqrt{\kappa_2} \exp(-\kappa_2 |z|), \ \
\epsilon_{02} = - \frac{\kappa^2_2}{2m},
\end{equation}
is solution of the equation:
\begin{equation}
\biggl[ -\frac{1}{2m} \biggl( \frac{d^2}{d y^2} \biggl) -
\frac{\kappa_2}{m} \ \delta(z) \biggl] \phi_{\epsilon_{02}}(z)=
\epsilon_{02} \ \phi_{\epsilon_{02}}(z).
\end{equation}
In the same way as for the amplitudes $A_m (H, \alpha)$ in
Eq.(13), it is possible to show that the amplitudes $B_l (H,
\alpha)$ in Eq.(18) satisfy the following equation:
\begin{eqnarray}
&B_l(H,\alpha)[\tilde \epsilon_2 - \epsilon_{02} + l \
\omega_c(\alpha)]
\nonumber\\
&+ B_{l+1}(H, \alpha) \ t_2 + B_{l-1}(H, \alpha) \ t_2 = 0 ,
\end{eqnarray}
where
\begin{equation}
t_2 =
\frac{\kappa_2}{m}\phi_{\epsilon_{02}}(0)\phi_{\epsilon_{02}}(\pm
c^*)= \frac{\kappa^2_2}{m} \exp(-\kappa_2 c^*).
\end{equation}
We pay attention that Eqs. (16) and (21) for the amplitudes, $A_m
(H, \alpha)$ and $B_l (H, \alpha)$, are similar, although the
cyclotron frequencies (6) have different signs in Eqs.(16) and
(21).

To solve Eqs.(16) and (21), we make use of the following
properties of the Bessel functions [28]:
\begin{equation}
z \ J_{n-1}(z) + z \ J_{n+1}(z) = 2n \ J_n(z),
\end{equation}
where $J_n(z)$ is the Bessel function of the n-th order. To this
end, we rewrite Eqs.(16) and (21) in the similar ways:
\begin{equation}
\frac{2t_1}{\omega_b(\alpha)} A_{m-1} +
\frac{2t_1}{\omega_b(\alpha)} A_{m+1} = 2 \biggl[m -
\frac{\epsilon_1 - \epsilon_0}{\omega_b(\alpha)} \biggl] A_m
\end{equation}
and
\begin{equation}
\frac{2t_2}{\omega_c(\alpha)} B_{l-1} +
\frac{2t_2}{\omega_c(\alpha)} B_{l+1} = 2 \biggl[m -
\frac{\epsilon_1 - \epsilon_0}{\omega_c(\alpha)} \biggl] B_l .
\end{equation}
Comparing Eqs.(24) and (25) with Eq.(23) for the Bessel functions,
we can conclude that Eqs.(24) and (25) have the following
solutions:
\begin{equation}
\tilde \epsilon_1 = \epsilon_{01} + n_1 \ \omega_b(\alpha), \ \ \
\tilde \epsilon_2 = \epsilon_{02} - n_2 \ \omega_c(\alpha)
\end{equation}
and
\begin{equation}
A_m(H, \alpha)=
J_{m-n_1}\biggl[\frac{2t_1}{\omega_b(\alpha)}\biggl],
\end{equation}
\begin{equation}
B_l(H, \alpha)=
J_{n_2-l}\biggl[\frac{2t_2}{\omega_c(\alpha)}\biggl],
\end{equation}
where $n_1$ and $n_2$ are some quantum numbers.

Therefore, electron wave functions (13) and (18) along ${\bf y}$
and ${\bf z}$ axes, respectively, can be rewritten as
\begin{equation}
\Psi_{\tilde \epsilon_1}(y) = \sum_{m=-\infty}^{+\infty} J_{m-n_1}
\biggl[\frac{2t_1}{\omega_b(\alpha)}\biggl] \
\phi_{\epsilon_{01}}(y - m b^*).
\end{equation}
and
\begin{equation}
\Psi_{\tilde \epsilon_2}(z) = \sum_{l=-\infty}^{+\infty} J_{n_2-l}
\biggl[\frac{2t_2}{\omega_c(\alpha)}\biggl] \
\phi_{\epsilon_{02}}(z - l c^*),
\end{equation}
where the total energy is
\begin{equation}
\epsilon = \epsilon_F + \epsilon_{01} + \epsilon_{02} + n_1 \
\omega_b(\alpha) - n_2 \ \omega_c(\alpha).
\end{equation}
From Eqs.(29) and (30), it directly follows that the electron wave
function is centered in the $n_1$-th conducting chains along ${\bf
y}$ axis and along the $n_2$-th conducting chains along ${\bf z}$
axis, respectively. Now, let us consider physical properties of
wave functions (29) and (30) for different directions and
strengths of a magnetic field. Suppose that a magnetic field is
applied far from the main crystallographic axis,
\begin{equation}
|\alpha| \sim 1, \ \ \ |\alpha - \pi/2| \sim 1,
\end{equation}
then at high enough magnetic fields,
\begin{equation}
H \gg \frac{2t_1 c}{e v_F b^*\cos \alpha}, \ \ \ H \gg \frac{2t_2
c}{e v_F c^* \sin \alpha},
\end{equation}
arguments of the Bessel functions in Eqs.(29) and (30) are small,
$2t_1 / \omega_b(\alpha) \ll 1$ and $2t_2 / \omega_c(\alpha) \ll
1$. As it follows, from the theory of the Bessel functions (see,
for example, [28]), the all Bessel functions with the exceptions
of that with $m=n_1$ and $l=n_2$ in Eqs.(29) and (30) are small
and, thus, the wave functions can be written in this case
approximately as
\begin{equation}
\Psi_{\tilde \epsilon_1}(y) \approx \phi_{\epsilon_{01}}(y - n_1
b^*).
\end{equation}
and
\begin{equation}
\Psi_{\tilde \epsilon_2}(z) \approx \phi_{\epsilon_{02}}(z - n_2
c^*).
\end{equation}
Note that the localization of the wave functions (34) and (35) on
one Q1D chain promotes non-Fermi-liquid properties. On the other
hand, if magnetic field direction is close enough to one of the
main crystallographic axes, corresponding in our case to
\begin{equation}
\alpha_1 =0, \ \ \ \alpha_2=\pi/2 ,
\end{equation}
then even in high magnetic fields (33) one of the Bessel functions
in electron wave functions (29) and (30) becomes delocalized [28].
The latter means that the Fermi-liquid properties have to restore
for angles (36). Therefore, we can conclude that at high enough
magnetic fields of the order of
\begin{equation}
H \geq H^* = \max \biggl[ \frac{2 \sqrt{2} t_1 c}{e v_F b^*}, \ \
\ \frac{2\sqrt{2} t_2 c}{e v_F c^* } \biggl]
\end{equation}
there have to be angular crossover (or phase transition) between
Fermi-liquid and non-Fermi-liquid states. [Note that in Eq.(37) we
have put $\alpha = \pi/4$.] It is important that wave functions
(29) and (30), derived in this paper by a full quantum mechanical
method, are similar to the quasi-classical wave functions (9) of
Ref.[26]. We pay also attention to the fact that the derived in
this paper equation for the corresponding critical field, $H^*$,
coincides with Eq.(12) from Ref.[26]. Therefore, we make a
statement that we justify hypothesis about the angular
Fermi-liquid - non-Fermi-liquid crossover (or phase transitions)
suggested in Ref.[26]. In particular, the corresponding critical
magnetic field for Q1D organic material (Per)$_2$Pt(mnt)$_2$ under
pressure can be estimated as $H^* \simeq 25 \ T$ [26].

In the paper, we have calculated fully quantum mechanical wave
function for the case, where a Q1D conductor is placed in a tilted
perpendicular magnetic field. We have evaluated energy in a
magnetic field with accuracy $\omega_b(\alpha) \sim \omega_c(\alpha)$
and disregarded only terms of the order of $\omega^2_b(\alpha)/\epsilon_F
\sim \omega^2_c(\alpha)/\epsilon_F$. As a result, we reproduced major
results of Ref.[26], where the so-called quasi-classical Peierls substitution
method was used. Our conclusion is that the Peierls substitution methods
is adequate not only for quasi-classical dimensional crossovers [4-14], where
the "sizes" of electron orbits are larger than inter-chain and inter-plane
distances, but also for quantum dimensional crossovers [21-27], where the
"sizes' of the orbits are less than inter-chain and inter-plane distances,
In some sense, in this paper we have validated previously obtained well-known
results [4-14], [21-27], and some others.

We are thankful to N.N. Bagmet, D. Graf, N.E. Hussey, and
J. Singleton for useful discussions.
 $^*$Also at: L.D. Landau
Institute for Theoretical Physics, RAS, 2 Kosygina Street, Moscow
117334, Russia.

\end{document}